\documentclass[runningheads]{cl2emult}

\usepackage{makeidx}  
\usepackage{graphicx} 
\usepackage{subeqnar} 
\usepackage{multicol} 
\usepackage{eso}      
\makeindex            

\def \thetas	{{\theta_{\rm S}}}
\def \thetae	{{\theta_{\rm E}}}
\def \thetai	{{\theta_{\rm I}}}
\def \dol	{{D_{\rm OL}}}
\def \dos	{{D_{\rm OS}}}
\def \dls	{{D_{\rm LS}}}
\def \abs	{ \hbox{ \vrule height .8em depth .4em width .6pt } \,} 
\def \lsim	{ \rlap{\lower .5ex \hbox{$\sim$} }{\raise .4ex \hbox{$<$} } }
\def \gsim	{ \rlap{\lower .5ex \hbox{$\sim$} }{\raise .4ex \hbox{$>$} } }

\def \msolar	{ \rm {M_{\odot}} }

\begin{document}

\title*{Planetary Microlensing: \protect\newline
Present Status and Long-term Goals}

\toctitle{Planetary Microlensing: Present Status and Long-term Goals}

\titlerunning{Planetary Microlensing: Present Status and Long-term Goals}

\author{Penny D. Sackett}

\authorrunning{Penny D. Sackett}

\institute{Kapteyn Astronomical Institute, 9700 AV Groningen, The Netherlands\\
psackett@astro.rug.nl}

\maketitle              


%


\section{Introduction}

Massive gravitational 
microlensing programs were begun about a decade ago as a 
means to search for compact baryonic dark matter in the Galaxy \cite{pac1986}, 
but before the first events were detected 
\cite{{alcock1993},{aubourg1993},{udalski1993}} the technique 
was also proposed as a means of detecting extra-solar planets 
in our Galaxy \cite{mandp1991}.  
\begin{figure}
\centering
\includegraphics[width=0.4\textwidth]{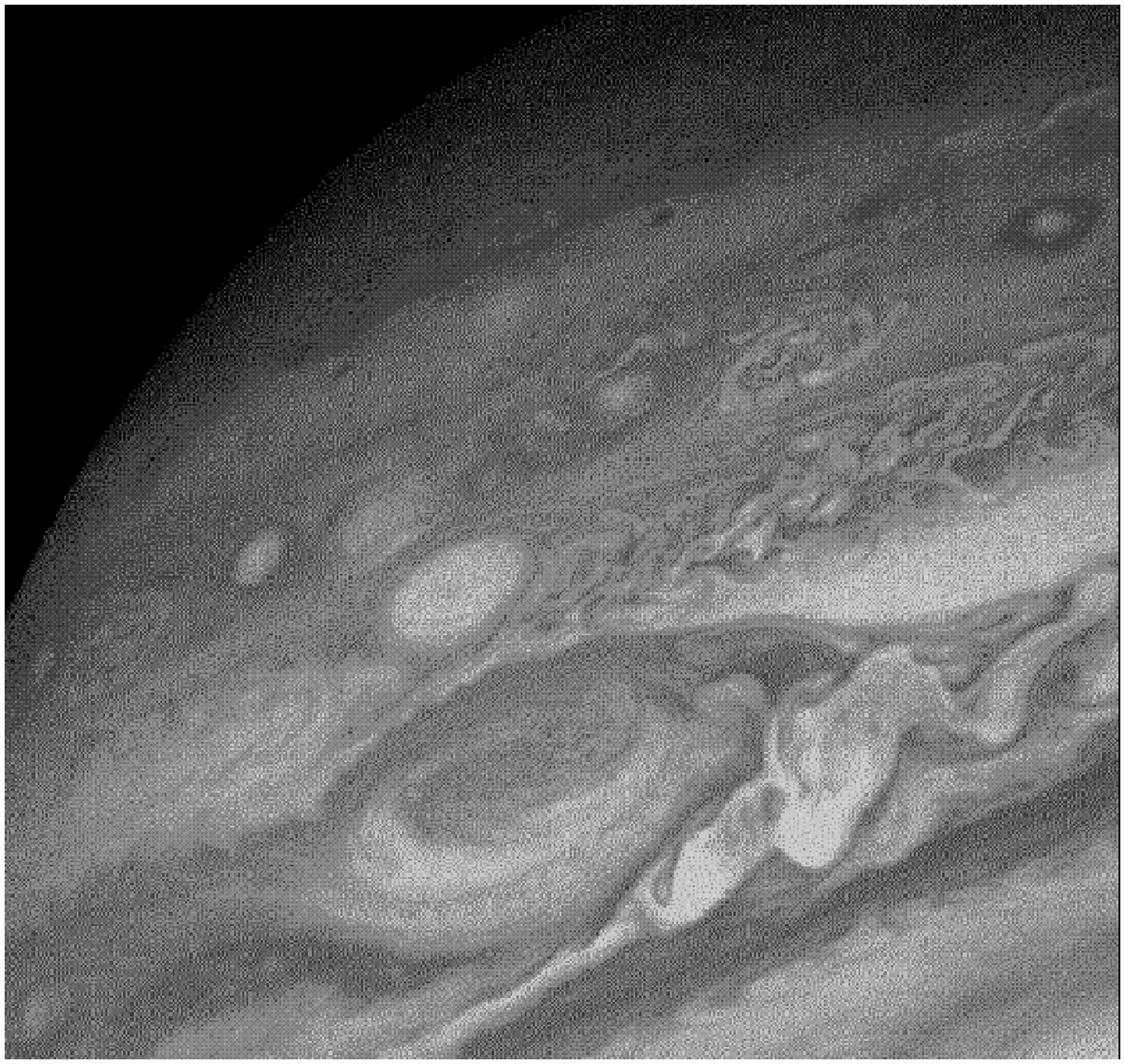}~~~~~
\includegraphics[width=0.377\textwidth]{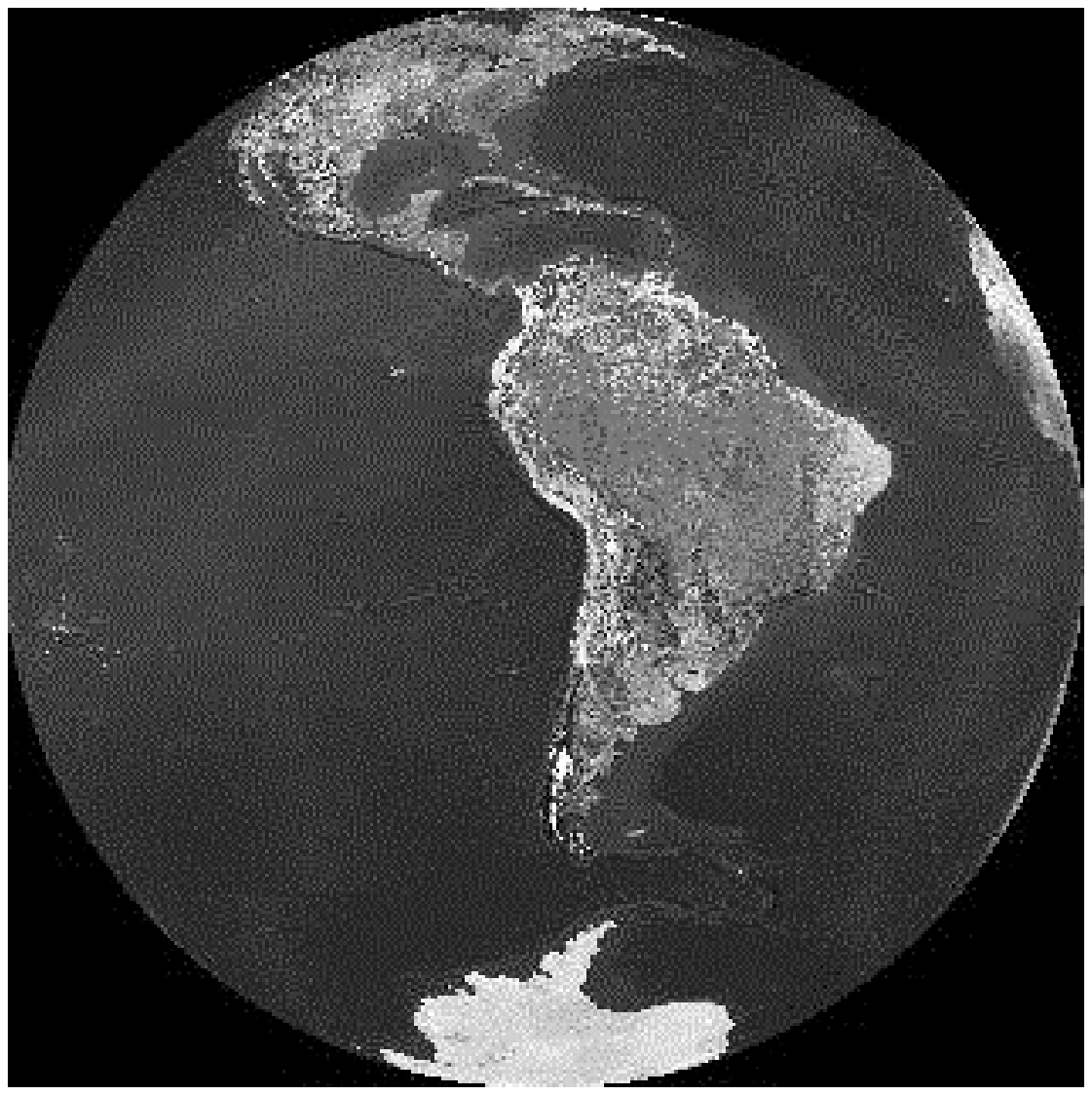}
\caption[]{Present microlensing planet detection programs are sensitive 
to planets similar to the one on the left; theoretical and observational capabilities must be increased by an order-of-magnitude before 
the long-term goal of reliably detecting planets similar to the one the 
right can be achieved.}
\label{planetpicies}
\end{figure}
\noindent
Current microlensing planet searches, which have been underway for four years, 
are sensitive to jovian-mass planets orbiting a few to several 
AU from their parent Galactic stars.  Within two years, sufficient data should 
be in hand to characterize or meaningfully constrain the frequency of 
massive planets in this range of parameter space, nicely 
complementing information about planets at smaller orbital radii 
now being provided by radial velocity searches.  
In principle, 
the technique could be pushed to smaller planetary masses, but only if a larger 
number of faint microlensed sources can be monitored with higher precision and 
temporal sampling.  The VST on Paranal, with spectroscopic follow-up with the 
VLT, may be the ideal instrument for such an ambitious program.


\section{Point lenses}

\begin{figure}
\includegraphics[width=0.6\textwidth]{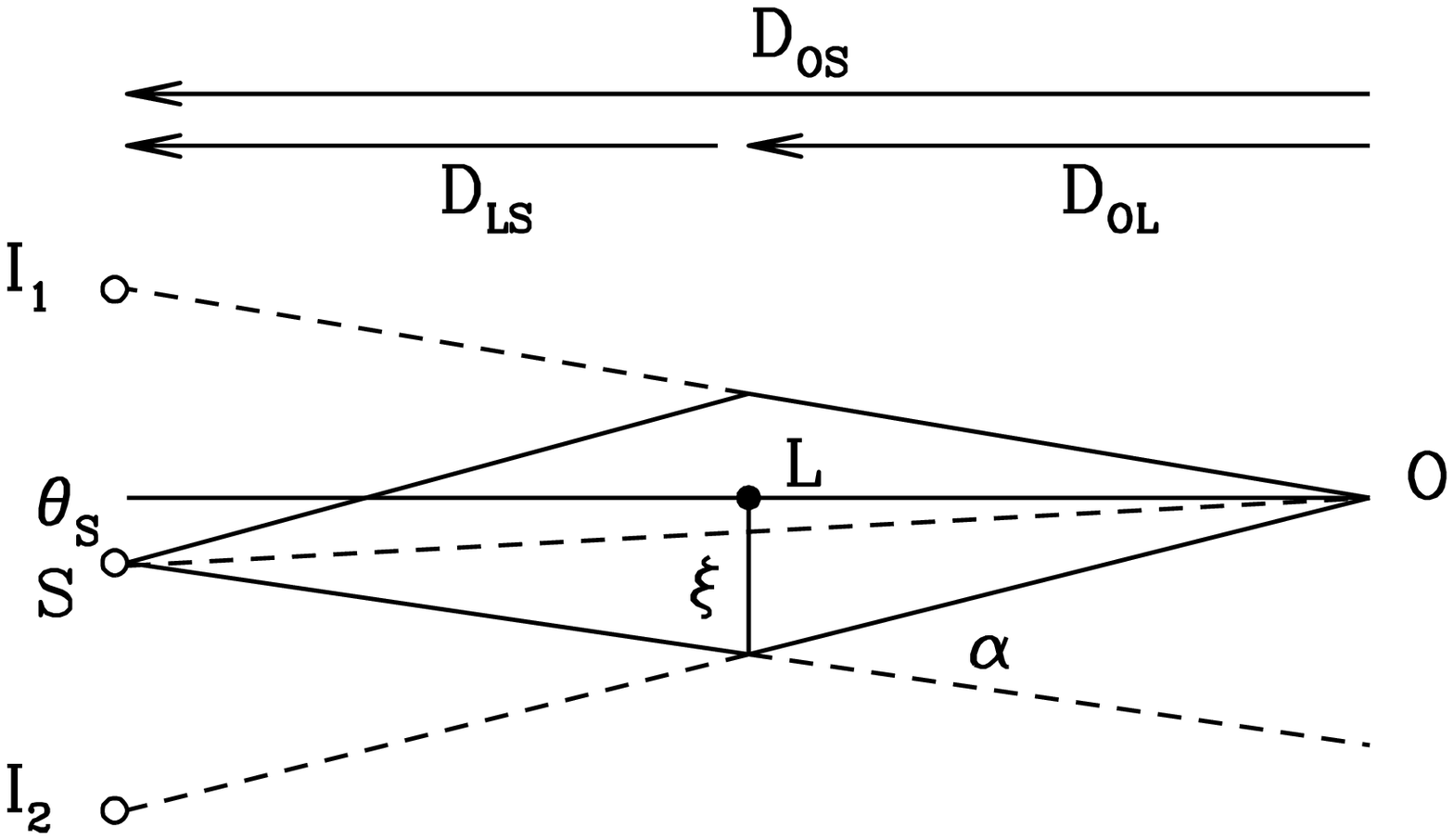}
\hglue-1.25cm\includegraphics[width=0.55\textwidth]{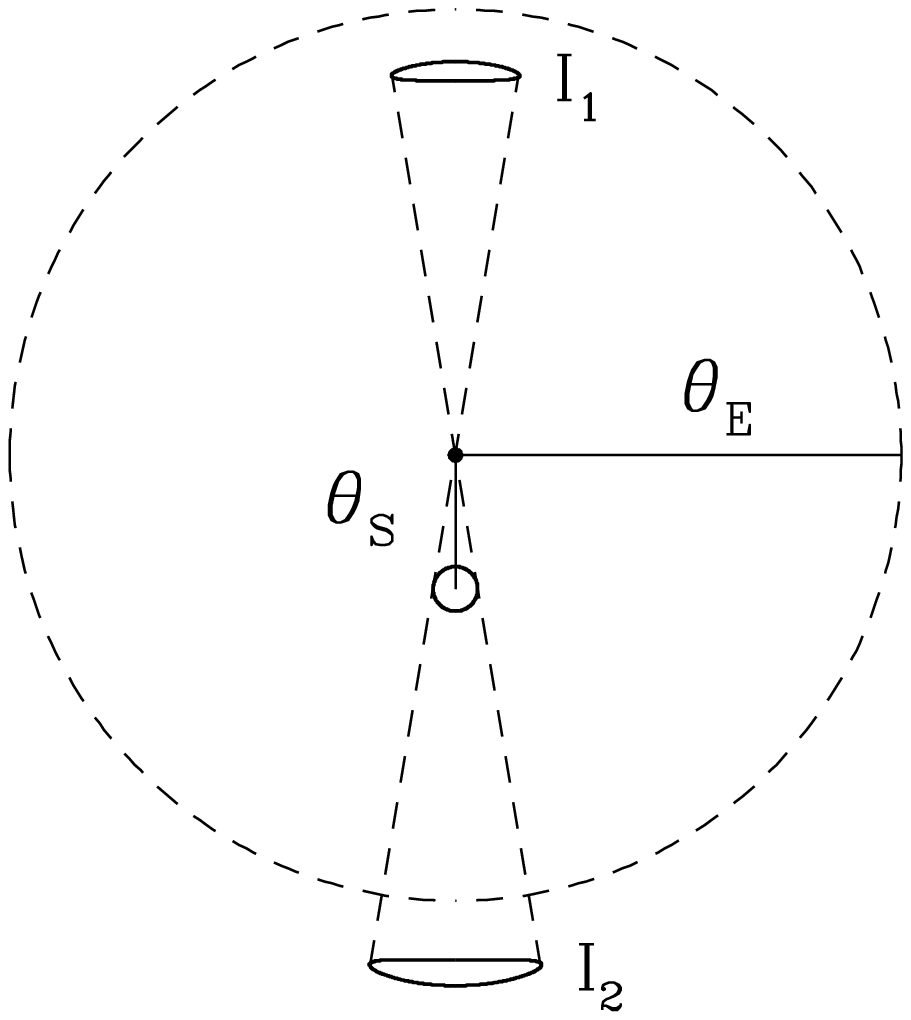}
\vglue -1.5cm
\caption[]{Point-lens microlensing geometry as seen from the side (left) 
and projected onto the sky (right).  The images I$_1$ and I$_2$ straddle 
the source position on the sky.}
\label{singlelensgeo}
\end{figure}

As light rays from a distant background source S pass a distance $\xi$ from a 
gravitational point lens L of mass $M$, they are bent by an angle 
$\alpha  = {4 \, G \, M}/{c^2 \, \xi}$. 
Simple geometric arguments reveal that the two resulting images 
have an angular separation on the sky 
$\Delta \theta_I = \sqrt{\thetas^2 + 4\, \thetae^2}$, where 
$\thetas$ is the angular distance between the lens and the observer-source  
sight line and $\thetae$ is the angular Einstein ring radius, a 
characteristic size for the lensing geometry defined by 
\begin{equation}
\thetae \equiv \sqrt{\frac{4 G M \dls}{c^2 \, \dol \, \dos}} ~~~. 
\label{ering}
\end{equation} 
\vskip 0.25cm\noindent
Here $\dol$, $\dos$, $\dls$ are the observer-lens, 
observer-source, and lens-source distances, respectively 
(Fig.~\ref{singlelensgeo}).  

\vglue-2.0cm

Since the specific intensity of each ray remains unchanged, 
the magnification of the images is just the ratio of the image area 
to the source area, which can be found by evaluating at the each image 
position the determinant of the Jacobian mapping $J$ describing the lensing 
coordinate transformation:
\begin{equation}
{\rm Magnification~Image~}_{i} 
= \left. \frac{1}{\abs \, det \, J \abs} \right|_{\, \thetai = \theta_{i}} 
= \left| \frac{\partial \, {\thetas}}{\partial \, {\thetai}} \right|^{-1}_{\, \thetai = \theta_{i}}~~. 
\label{mageq}
\end{equation} 
\noindent 
The largest magnifications occur when the determinant is near zero. 
For a point lens this occurs when $\thetas \approx 0$ and the source lies 
almost directly behind the lens; the images then lie 
close to the Einstein ring $\thetae$. 
For lenses with stellar masses and distances typical for stars 
within the Milky Way, $\thetae \sim 1\,$mas, so that whenever the images 
are significantly magnified, they are too close together to be 
resolved by traditional imaging.  Only the combined magnification $A$ of 
both images can be measured.  For a point-lens, Eq.~\ref{mageq} 
can be used to show that $ A = (u^2+2)/(u \sqrt{u^2+4}) $, 
where $u \equiv \thetas/\thetae$ is the instantaneous 
source-lens separation in units of the angular Einstein radius.  

\begin{figure}
\vglue-0.75cm
\hglue-1cm\includegraphics[width=0.65\textwidth]{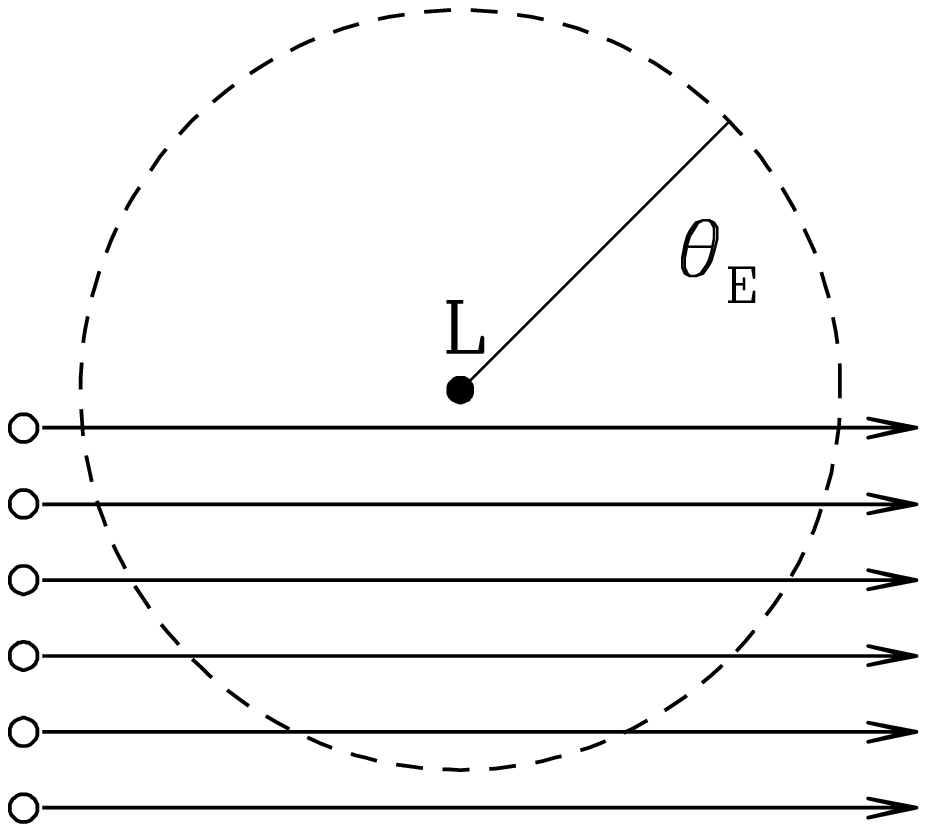}
\hglue-2.0cm\includegraphics[width=0.6\textwidth]{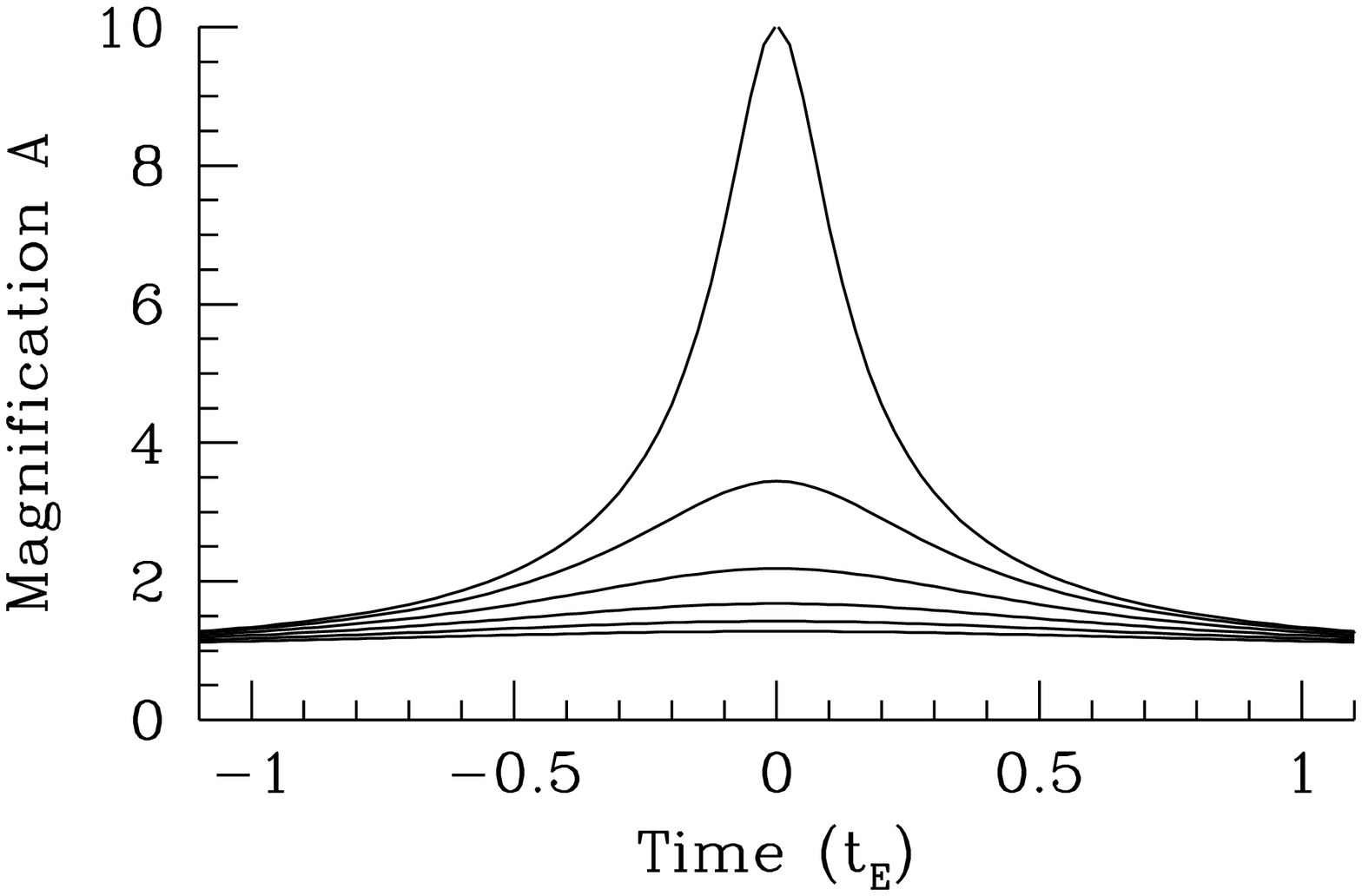}
\vglue -2.25cm
\caption[]{Sample source trajectories (left) and the resulting light curves 
(right).}
\vglue -0.25cm
\label{singlelenslcs}
\end{figure}

Because the source, lens, and observer are in relative motion, $u$ 
is a function of time. When $u$ is at its minimum, 
the observed light curve of the microlensing event undergoes its maximum 
magnification.  Microlensing events for which the source has a small 
minimum impact parameter $u_{\rm MIN}$ will have the largest peak 
amplifications (Fig.~\ref{singlelenslcs}).  
The characteristic time 
$t_{\rm E} \equiv \thetae \dol / v_\perp$ is the time taken by the 
lens, moving at speed $v_\perp$ across the sight line to the source, 
to travel one Einstein radius, and is a several days  
to months for most Galactic microlensing events.


\section{Binary Lenses}

\begin{figure}
\vglue -1.0cm
\hglue0.25cm\includegraphics[width=0.45\textwidth]{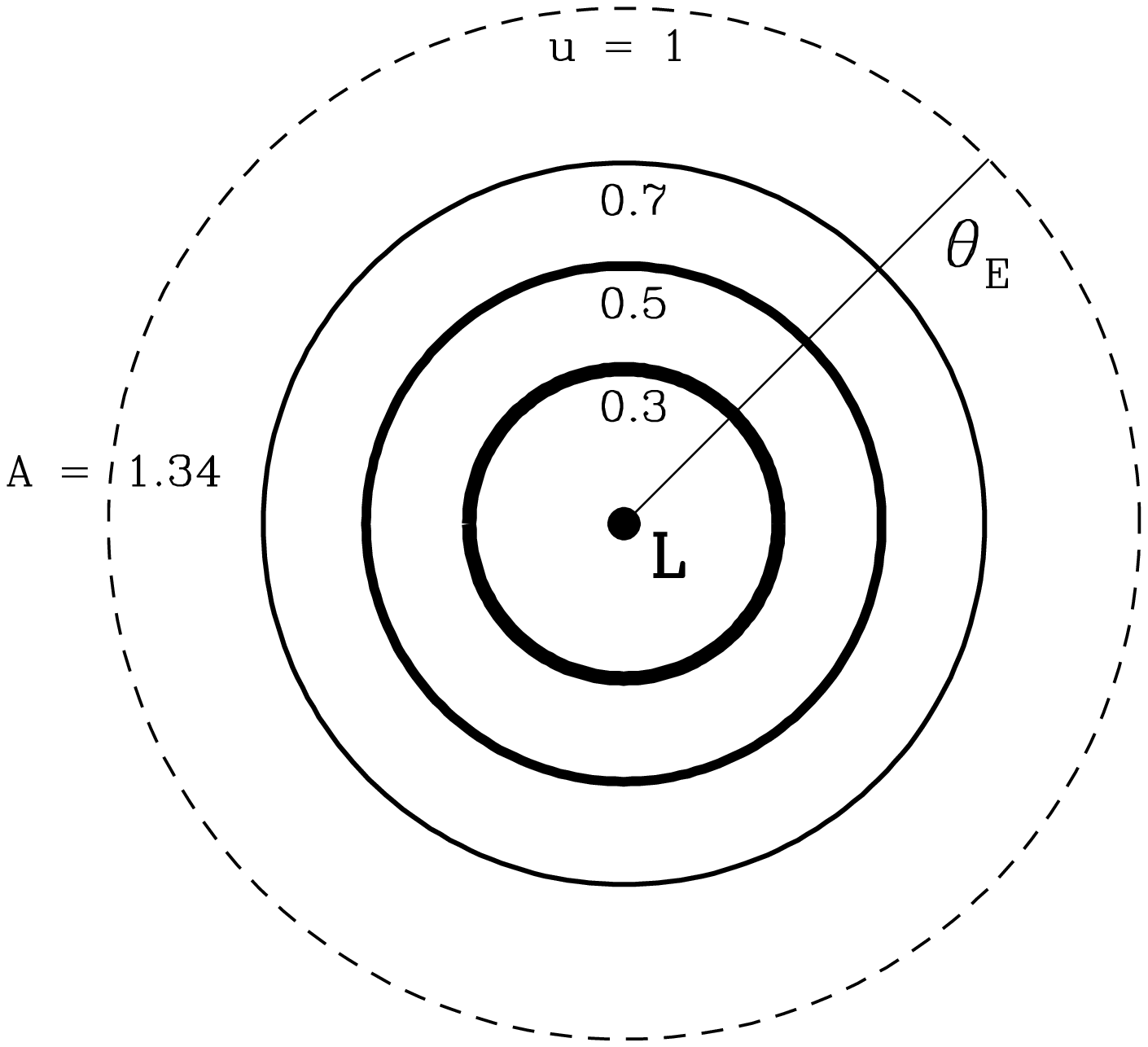}
\vglue -8.0cm
\hglue 7cm\includegraphics[width=1.55\textwidth]{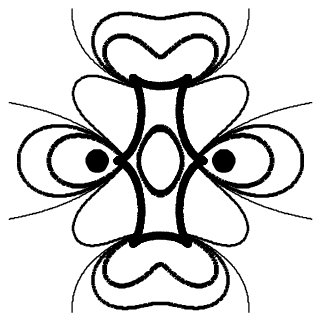}
\vglue -11.20cm
\caption[]{Single lens magnification map (left) and a binary lens 
fractional deviation map (right).  
The binary components (heavy dots) are 
separated by 1~$\thetae$ and have a total mass equal to the single lens.  
Regions where the binary magnification 
is depressed or increased by 1\% and 5\% relative to that of the single lens 
are marked by light and bolder contours, respectively. 
The very bold contour is the caustic.}
\label{magpatterns}
\end{figure}

Point lens light curves are symmetric because they are 
one-dimensional cuts through two-dimensional circularly-symmetric 
magnification patterns on the sky.  
Double lenses can generate more complicated light curves because 
this symmetry is broken.  In particular, the loci of points in the 
source plane for which $\abs \, det \, J \abs = 0$ have quite complicated structure (Fig.~\ref{magpatterns}).  

These loci --- called {\it caustics\/}  
--- mark positions at which point sources 
would experience infinite image magnification. 
The observed flux remains finite as a caustic crosses a real source; 
integration over the source size $\theta_*$ is required to compute 
the observed total magnification.  Sources passing near binary-lens caustics 
will exhibit light curves that deviate strongly from those of single lenses.  
The smaller the source and the closer the caustic approach, the larger 
the deviation will be.  Unlike single lens curves, the shape of a  
binary light curve depends on the angle of the 
source trajectory through the magnification pattern.  

Any static, unblended binary light curve is described by eight parameters: 
four quantities relevant to single lenses  
($u_{\rm MIN}$, $t_{\rm E}$, time at peak $t_0$, and baseline flux $F_0$) 
and four additional parameters, namely, 
the binary mass ratio $q \equiv m_1/m_2$, 
the instantaneous angular separation $b$ of the components in 
units of $\thetae$, 
the ratio $\rho_* \equiv \theta_*/\thetae$ of source radius to 
Einstein radius, and the angle $\phi$ between the 
source trajectory and the binary axis.


\section{Planetary Microlensing}

\begin{figure}
\vglue -0.5cm
\centering
\hglue-0.4cm\includegraphics[width=0.75\textwidth]{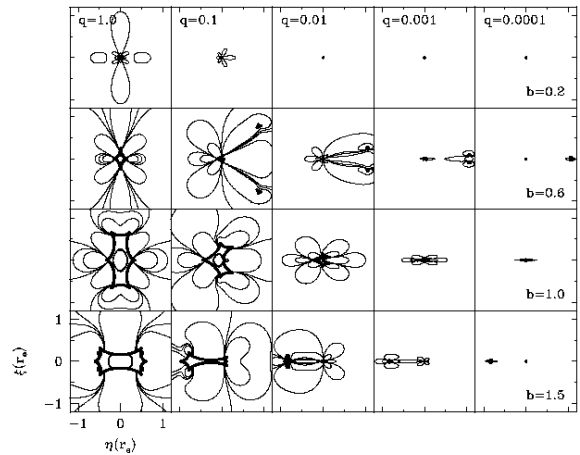}
\vglue -2.6cm
\caption[]{Fractional deviation patterns for binary lenses with 
different mass ratios $q$ and separations $b$. Contours are the same as 
in Fig.~\ref{magpatterns}; heavy lines are caustics.  
A ``super-jupiter'' with mass equal to 10 $M_{\rm J}$ orbiting 
a late M dwarf would have $q \approx 0.1$; $q\ \lsim\ 0.01$ would 
almost certainly correspond to a planetary rather than stellar binary system.  
Adapted from Gaudi \& Sackett 1999.}
\label{binarycausticpanel}
\end{figure}

A lens orbited by one or more planets is a multiple lens and thus 
may have a magnification pattern that differs measurably 
from that of a single lens.  Measuring and characterizing light curve 
deviations induced by planetary lensing companions is the goal of  
microlensing planet detection programs.
The number, size and relative positions of 
the caustics depend on the mass ratio $q$ and separation $b$ of the 
double lens (Fig.~\ref{binarycausticpanel}).  

\begin{figure}
\vglue -7.85cm
\hglue-0.2cm\includegraphics[width=1.05\textwidth]{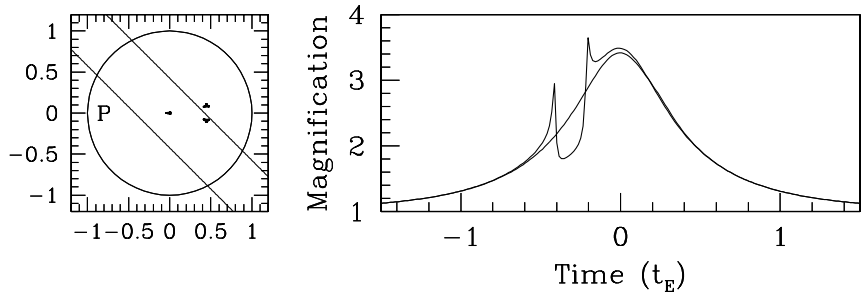}
\vglue -0.75cm
\caption[]{Light curves (right) generated by two source trajectories 
(left) through the same magnification pattern due to a stellar lens (center)  
orbited by a planet (``P'') with $q=0.003$ at a separation 
$b = 0.8 \thetae$.  One light curve reveals the planet; the other does not. 
Note the three tiny caustics.  The central caustic is 
always located close to the primary. The planetary caustics 
do not coincide with the planet, but their number and location do depend 
on the planet's position and mass ratio.}
\vskip -0.5cm
\label{sampleplanetlc}
\end{figure}

The duration, amplitude and placement of a planetary anomaly atop 
an otherwise normal microlensing light curve depends on the mass 
ratio $q = m_p/M$, instantaneous separation $b$, and the source trajectory. 
Well-sampled light curves thus allow the determination of $q$ and $b$  
if an anomaly is detected, but not all source trajectories will 
generate a detectable anomaly even if the lens has a massive planet in 
its lensing zone (Fig.~\ref{sampleplanetlc}).  
     
Each planet in a lensing system will generate isolated planetary caustics 
that will influence the overall magnification pattern in a nearly independent 
manner (Fig.~\ref{sampleplanetlc}); 
for most events, a given planet will be detected only if the 
source passes near one of these caustics.  The central caustic, 
on the other hand, is affected by any planet in the system \cite{gands1998}, 
so that high magnification events ---  
in which the source always passes close to the primary lens --- 
are sensitive to the presence of {\it multiple planets\/} \cite{gns1998}, 
though attaching a unique set of multiple planets to a given deviation 
will be difficult due to the increased complexity of the 
caustic structure.


\section{Capabilities of Current Microlensing Planet Searches}

Light curve morphologies are quite varied \cite{wambs1997}, 
but broadly speaking both the caustic cross section 
presented to a (small) source and the duration of a planetary anomaly 
are proportional to $\sqrt{q}$ \cite{dom1999}.  
Larger planets are easier to detect 
both because it is more likely that the source will pass 
near a caustic and because the perturbation lasts longer.
For Galactic stellar lenses, jovian-mass planets are  
expected to have durations of 1-3 days; terrestrial-mass 
anomalies will last only a few to several hours.   
As Fig.~\ref{binarycausticpanel} illustrates, stellar binaries
($1\ \lsim\ q\ \lsim\ 0.1$) in the Lensing Zone nearly always create 
perturbations larger than $1\%$, regardless of 
source trajectory.  Planetary companions ($q\ \lsim\ 0.01$)  
can escape detection much more easily since it likely that 
the source will not cross any of the deviation contours that are 
above the photometric noise.  
The efficiency with which planets can be detected in a 
given microlensing data set will depend therefore not only 
on the separation and mass ratio of the planet, but also on 
the temporal sampling and photometric precision.  Studies of simulated 
data sets using different detection criteria and assuming  
light curves monitored continuously with the best precision currently 
possible in crowded fields have estimated that  
planets like our own Jupiter ($q = 10^{-3}$) may be detectable with 15$-$40\% efficiency \cite{{gandl1992},{bandf1993}}.  Efficiences for observed data 
sets with erratic photometric quality and sampling 
must be computed separately for each light curve \cite{gands1999}.  
Detection efficiencies of most light curves observed today 
are effectively zero for earth-mass ($q = 3 \times 10^{-6}$) planets.

Planets too close to their parent lenses on the sky 
will closely resemble a single combined lens; planets too 
widely separated will behave like isolated single lenses.  
Planet-star systems with 
separations $0.6\ \lsim\ b\ \lsim\ 1.6$ 
--- that is, planets inside the so-called {\it Lensing Zone\/} --- 
generate the most prominent binary caustic structure inside 
the Einstein ring $\thetae$ of the primary (Fig.~\ref{binarycausticpanel}). 
Since microlensing events are seldom alerted and monitored 
unless the source is inside $\thetae$  
(i.e, $u < 1$), current surveys are most 
sensitive to planets in the Lensing Zone.  
Planets outside this zone could be detected by microlensing 
if the light curve is monitored for source positions outside $\thetae$ in 
order to have sensitivity to distant, outer planetary caustics \cite{dands1999} 
or, a in very high amplification events which bring the source close to 
the central caustic generated by all planets \cite{{gands1998},{gns1998}}. 
A $1\ \msolar$ lens positioned halfway to source stars in the Galactic 
Bulge (the location of the overwhelming majority of events) 
has a physical Einstein ring radius of 4~AU.  
Most lenses will be somewhat less massive and closer to the Bulge, 
yielding Lensing Zones between 1 and 6~AU.  
Depending on their orbital inclination, planets with larger 
orbital separations may be brought into this zone for a 
certain fraction of their orbit.  
In sum,
current microlensing searches are sensitive to massive ($q > 10^{-4}$) planets 
orbiting 1 -- 10~AU from their parent stars, a region rich with planets 
in our own Solar System.  

Three teams, PLANET \cite{albrow1998}, MPS, and MOA, 
now routinely use longitudinally distributed networks of southern 1m telescopes
to monitor events discovered by microlensing surveys \cite{{alcock1993},{aubourg1993},{udalski1993}}, with  
the detection of planetary anomalies as one of their primary goals.  
Of the three, PLANET currently enjoys the most extensive network 
of semi-dedicated telescopes, and performs intensive, nearly continuous 
(every ~2 hours, weather permitting) photometric monitoring of 
several events per night with $\sim 1 - 5$\% precision \cite{albrow1998}.  

\begin{figure}
\vglue -1.00cm
\hglue 1.0cm\includegraphics[width=0.8\textwidth]{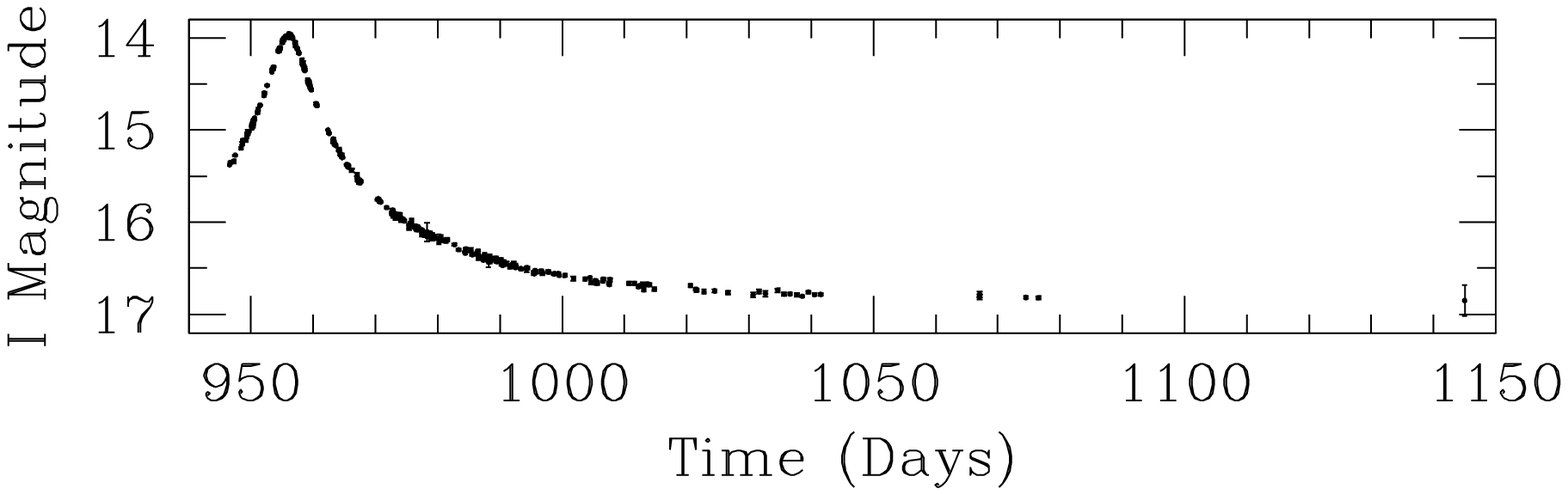}
\vglue -6.4cm
\hglue 2.0cm\includegraphics[width=0.6\textwidth]{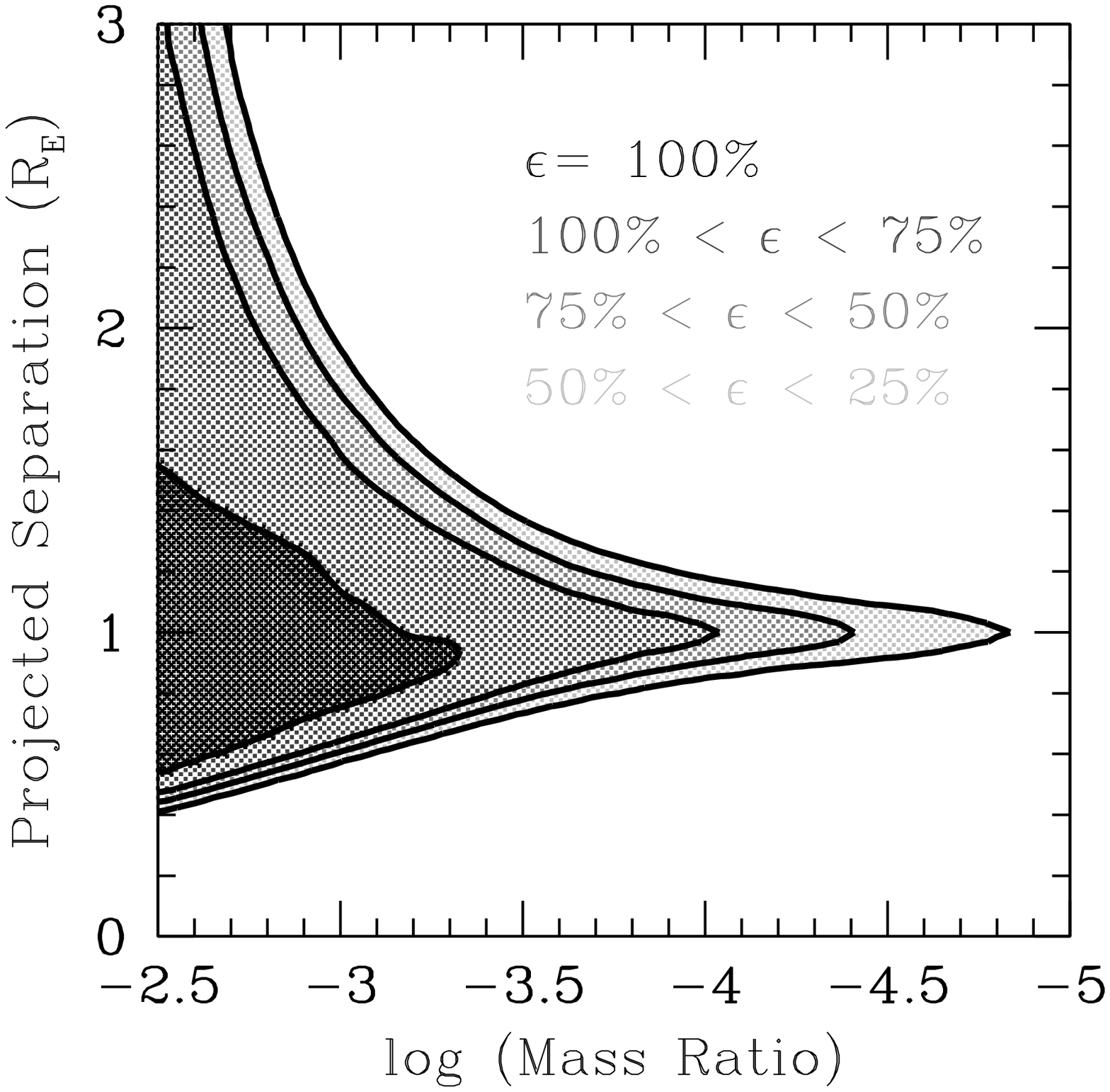}
\vglue -0.4cm
\caption[]{{\it Top:\/}
The PLANET collaboration I band light curve containing $\sim$400 points for 
event OGLE~98-BLG-14 does not differ obviously from that of a single lens.  
{\it Bottom:\/}  Exclusion probability contours for planetary companions 
of given mass ratio and instantaneous separation (in Einstein radii) 
orbiting this microlens \cite{{gaudiO14},{albrowO14}}.  
Within the solid black region, 
the best model with a planet differs from that without a planet 
by $\Delta \chi^2 > 100$ for all source trajectories.}
\vglue -0.75cm
\label{O14fig}
\end{figure}

No convincing planetary signal has yet been detected, though not all 
data have been analyzed thoroughly.  However, by comparing models with and 
without planets for all possible source trajectories \cite{gands1999}, 
the presence of massive planets within certain zones of angular  
separation from their parent lenses can be {\it ruled out\/} in 
very well sampled, high magnification events \cite{{gaudiO14},{albrowO14}} (Fig.~\ref{O14fig}).  The accumulation 
of many such events will lead in the next few years 
to the detection of planets like our 
own Jupiter --- or to meaningful constraints on the abundance of such 
planets in the Galaxy (Fig.~\ref{compare}). 

\section{Planetary Microlensing in the VLT Era}

Small planets are difficult to detect with any method; microlensing 
is no exception.  Earth-mass planets are especially elusive because 
the angular size of the planetary caustics are smaller than a giant star 
in the Bulge.  Thus, even if an earth-mass caustic directly 
transits a background giant, only a piece of source 
is significantly magnified; source resolution greatly dilutes the signal 
\cite{bandr1996}.  
In order to detect such small planets, 
several obstacles associated with earth-mass 
microlensing anomalies must be overcome simultaneously, namely their:
(1) rarity, (2) short duration, (3) typically small amplitude, 
and (4) near invisibility against giant sources.  A successful program 
would thus need to monitor hundreds of dwarf or turn-off stars 
undergo microlensing in the Bulge, and to do so frequently and with high 
precision.

The 2.5m VST equipped with OmegaCAM, a 16K $\times$ 16K CCD detector spanning 
a 1 degree field of view, is expected to see first light on Paranal in 2001 
as a survey telescope for the VLT.  The excellent weather and median 
seeing (0$''$.65), large field, and possibility of immediate VLT follow-up 
could make the VST the most formidable microlensing machine of its era;  
{\it simultaneous detection and monitoring of 10 -- 20 on-going 
microlensing events in every bulge field\/} would be possible.  
Detailed observing strategies and simulations for the VST remain 
to be worked out, but first estimates are encouraging.  
Assuming that 1\% photometry on V=20 (I=19) turn-off stars can be achieved 
with 4-5 minute integrations in these very dense fields, and allowing for 
30-50\% lost time due to weather or poor seeing, continuous 
observations with the VST during ``bulge season'' could yield as many as 
$\sim$20 jovian and $\sim$2 terrestrial-mass planets 
per year --- if every lens has one of each sort of planet in its lensing zone 
\cite{{sackett1997},{peale1997}}.  Since this is unlikely to be 
the case, detected numbers will be smaller, but it is only by  
measuring how much smaller that we can determine  
the abundance of such planets in our Galaxy.

Since microlensing ``selects'' lensing stars by mass, not luminosity, 
very distant and dim stars can be probed for planetary systems.  
Furthermore, the selection 
is a weak function of mass, so all types of Galactic stars can be studied.   
Unfortunately, because the parent star is unseen, its mass and distance 
are unknown: generally stellar lenses are too close (mas) to the 
source to be resolved with normal imaging 
and too faint to be detected in a combined spectrum.  
Lens mass $M$ and distance $\dol$ are 
the scaling parameters that are required to translate the 
planetary mass ratio $q$ into a planetary mass $m_p$ and the normalized 
instantaneous separation $b$ into 
physical units such as AU (see Eq.~\ref{ering}).  
Without knowledge of the lens mass and distance,  
ensembles of events must be fit with reasonable Galactic models 
to derive statistical estimates for these quantities.

On the other hand, 8-10m telescopes offer the first hope to 
spectrally type the otherwise unseen microlenses so that their distance 
and mass can be determined directly.  Because the source and lens 
star are likely to differ in both spectral type and radial 
velocity, high resolution spectra 
($\lambda/\Delta \lambda = 40000$) with large apertures are 
expected to detect the lens signal in composite  
spectra, even for lens-source contrasts of 4 magnitudes --- 
allowing dwarf lenses to be discerned \cite{mrl1998}.  
In order to make best use of the potential for scientific gain, 
the microlensing events should be identified in real time at the VST 
so that VLT spectra can be taken both at peak, when the magnified source 
spectral energy distribution will dominate, and nearer baseline, when 
the composite spectrum will reflect the unlensed fraction of light 
from source and lens.  
This ambitious goal of direct lens detection would 
open a new era for microlensing, and is a task that the VLT is 
especially well suited to tackle due to its flexible instrumentation 
and scheduling, proximity to the VST, 
and the excellent seeing conditions on Paranal.
The speed with which VLT instrumentation 
can be made ready and the availability of service observing will make 
it ideal for other target-of-opportunity science as well, 
such as the monitoring of (1) caustic crossings for proper motion 
and limb-darkening measurements, (2) supernovae, and (3) 
gamma ray burst optical counterparts. 

\begin{figure}
\vglue -1.7cm
\hglue 1.65cm\includegraphics[angle=90,width=8.95cm]{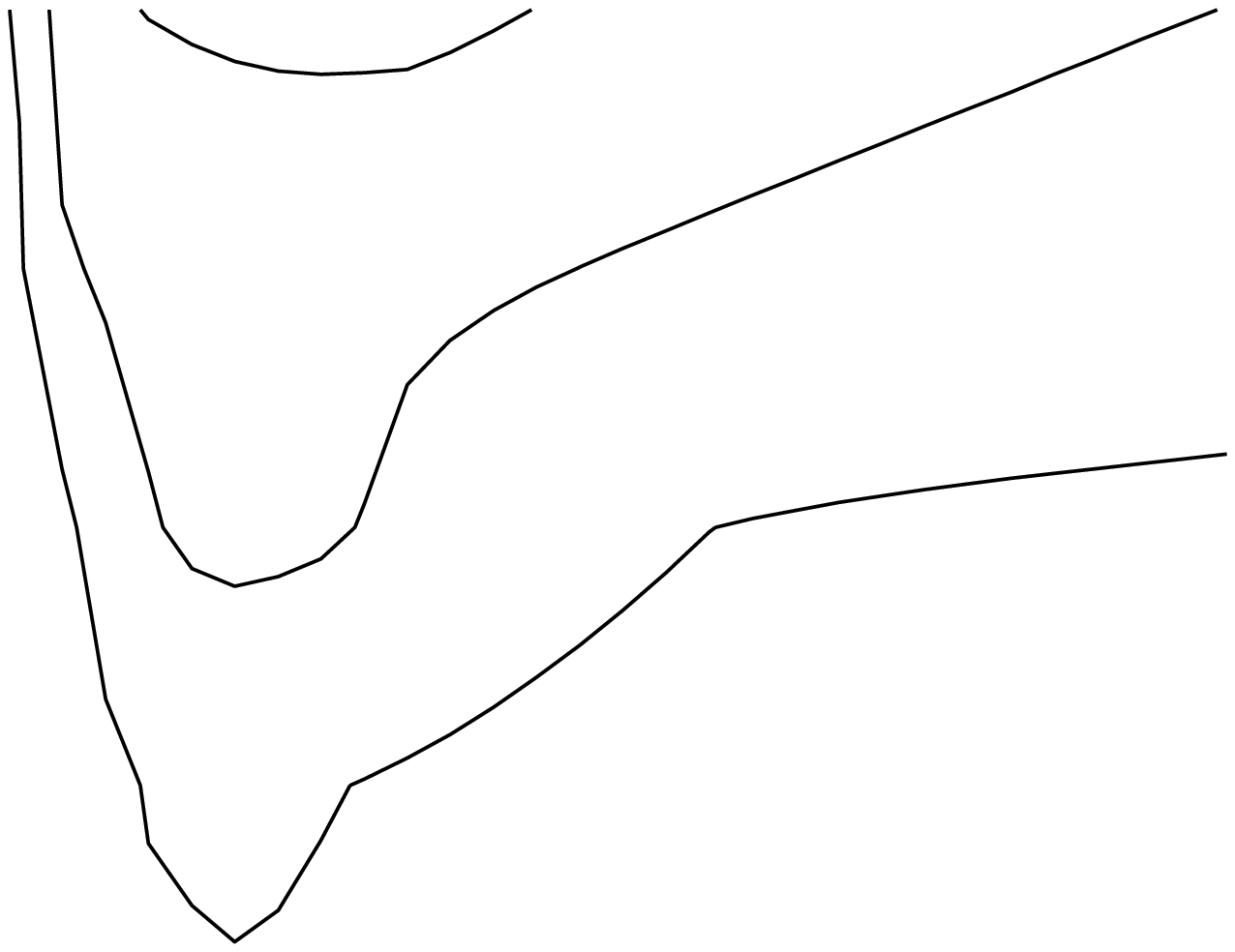}
\vglue -7.7cm
\hglue 2.45cm\includegraphics[width=7.05cm]{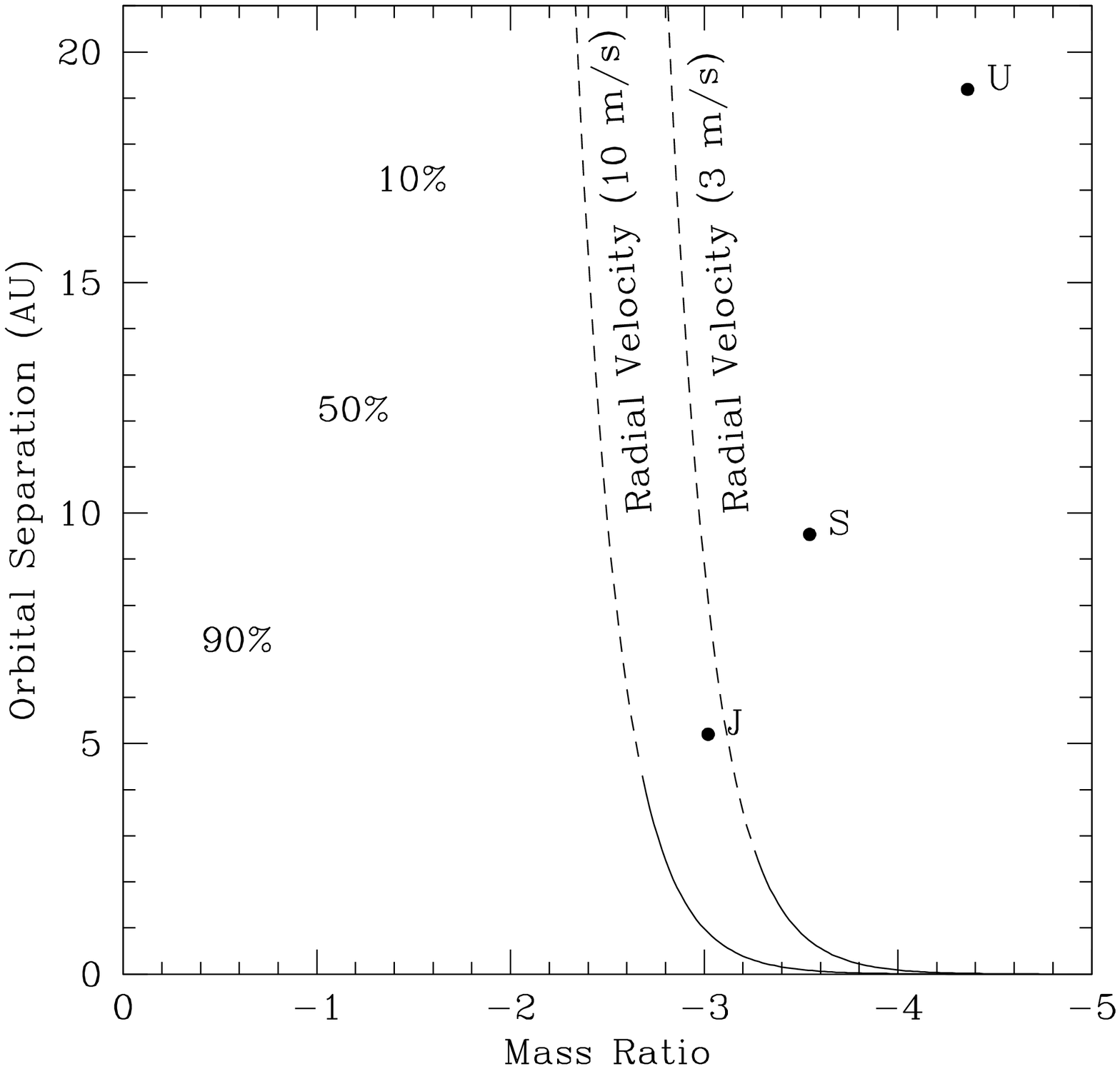}
\caption[]{Estimated detection efficiency contours of 10, 50, and 90\% 
for current microlensing searches for planets of given mass ratio and true 
orbital separation $a$ in units of AU. 
(The Einstein ring radius is taken to be 3.5~AU, appropriate 
to a solar type Bulge lens.) Also shown are expectations 
for a radial velocity planet search running for 5 and 10 years (solid lines) 
and requiring 3-$\sigma$ detections at limiting sensitivities of 
3 and 10 m s$^{-1}$, respectively.  
Positions of Jupiter, Saturn 
and Uranus are shown for reference. Adapted from Sackett (1999).} 
\vglue -0.25cm
\label{compare}
\end{figure}

As Fig.~\ref{compare} makes evident, if a sufficient number of events 
can be monitored sufficiently well, current microlensing searches 
will contribute to our knowledge of jovian mass planets orbiting 
with true orbital separations comparable to and somewhat larger than 
that of our own Jupiter, complementing current radial velocity searches 
which are sensitive to (and finding!) 
jovian planets at smaller orbital radii ($\lsim 3 \,$AU).  
The VLT (and VST) era may see a widening 
of the parameter space to which microlensing can contribute 
valuable information, especially by pushing the technique forward to 
lower masses while allowing an improved characterization of detected 
planetary systems through better photometry and --- possibly --- 
direct spectroscopic microlens detection.


\subsection*{Acknowledgements}

It is a pleasure to thank my colleagues in the PLANET collaboration 
for permission to display our results for OGLE~98-BLG-14 prior to publication, 
and Scott Gaudi for assistance in preparation of Fig.~8. 


\clearpage
\addcontentsline{toc}{section}{Index}
\flushbottom
\printindex

\end{document}